\documentclass[english]{article}
\usepackage[T1]{fontenc}
\usepackage[latin9]{inputenc}
\usepackage{amsmath}
\usepackage{amssymb}
\usepackage{esint}

\makeatletter
\usepackage{babel}

\begin{document}
\begin{titlepage}
\thispagestyle{empty}

\bigskip

\begin{center}
\noindent{\Large \textbf
{Quantum Kalb-Ramond Field in D-dimensional de Sitter Spacetimes}}\\

\vspace{0,5cm}

\noindent{G. Alencar ${}^{a}$\footnote{e-mail: geovamaciel@gmail.com }, I. Guedes ${}^{a}$, R. R. Landim ${}^{a}$ and R.N. Costa Filho ${}^{a}$}

\vspace{0,5cm}

 {\it ${}^a$Departamento de F\'{\i}sica, Universidade Federal do Cear\'{a}-
Caixa Postal 6030, Campus do Pici, 60455-760, Fortaleza, Cear\'{a}, Brazil. 
 }

\end{center}

\vspace{0.3cm}

\begin{abstract}

In this work we investigate the quantum theory of the Kalb-Ramond fields
propagating in $D-$dimensional de Sitter spacetimes using the dynamic invariant method
developed by Lewis and Riesenfeld [J. Math. Phys. 10, 1458 (1969)]  to obtain the
solution of the time-dependent Schr\"odinger equation. The wave function is written in
terms of a $c-$number quantity satisfying of the Milne-Pinney equation, whose solution
can be expressed in terms of two independent solutions of the respective equation of
motion. We obtain the exact solution for the quantum Kalb-Ramond field in the de
Sitter background and discuss its relation with the Cremmer-Scherk-Kalb-Ramond
model.

\end{abstract}
\end{titlepage}

\section{Introduction}

Although the Quantum Field Theory in the Minkowski space is well developed,
there still is a question related to the behavior of fields in different backgrounds in
cosmological scales. The string theory is likely the best tool used to solve the problem
of quantum gravity \cite{Polchinski:1998rq,Polchinski:1998rr,Berkovits:2000fe}, even though other problems can be analyzed in the light of
fields propagating in curved spacetimes \cite{Birrell:1982ix}. Several interesting results were found in
time-dependent backgrounds, namely: (i) black hole evaporation \cite{Hawking:1974rv}, (ii) the Unruh \cite{Crispino:2007eb}
and Casimir \cite{Saharian:2009ii} effects and (iii) the particle creation \cite{Parker:1968mv,Parker:1969au,Parker:1971pt}.

The quantum effects of a massive scalar field in the de Sitter spacetime
from a Schr\"odinger-picture point of view has been investigated in Ref.
\cite{pedrosa1}.The light propagation through
time-dependent dielectric linear media in the absence of free charges
and in a curved spacetime from a classical and a quantum point of
view was presented in Ref. \cite{pedrosa2}. 
Scalar fields have also been investigated in Ref. \cite{MM}, where
the coherent and squeezed states were investigated. 

In these previous works (and in Refs. \cite{Pedrosa_97,Pedrosa_05}), the authors used the Lewis and Riesenfeld method \cite{Lewis:1968tm}
to study the quantization of fields propagating in $D=4$. Recently, we
also have used the Lewis and Riesenfeld method to study the quantization of both scalar
\cite{Alencar:2011an} and electromagnetic fields \cite{Alencar:2011nw} in arbitrary D-dimensions. In Ref. \cite{Alencar:2011an} we found
that the Bunch-Davies thermal bath depends on the choice of $D$ and the conformal
parameter $\xi$, which is important in extra-dimension physics, e.g. in the Randall-Sundrum
models. In Ref. \cite{Alencar:2011nw} we found that only for $D=4$ there is no thermal bath as expected.

In this paper we study the problem of quantizing the Kalb-Ramond field in the
$D-$dimensional de Sitter spacetime. The Kalb-Ramond field has been studied in $D=4$ as
a description of an axion field \cite{Gasperini:1998bm,Giovannini:1998ig}. In extra dimensional scenarios, it has been used
to describe torsion \cite{Mukhopadhyaya:2009gp} and brane interactions \cite{Barone:2010zi}. The interest in the concept of extra
dimension has been renewed by String Theory and the Randall-Sundrum models \cite{Randall:1999vf,Randall:1999ee}. The effective action for the
standard model and the localization of fields within of an extra dimensional scenario
have been considered in Refs. \cite{Csaki:2002gy} and \cite{Landim:2011ki,Landim:2011ts}, respectively. The Kalb-Ramond field emerges
naturally in the spectrum of the closed string \cite{Polchinski:1998rq,Polchinski:1998rr} and its presence may induce optical
activity in the visible brane \cite{Kar:2002xa}.

This paper is organized as follows. In Sec. II the correspondence between the
Kalb-Ramond field placed in a de Sitter background and a time-dependent harmonic
oscillator is obtained. In Sec. III, we use the Lewis and Riesenfeld \cite{Lewis:1968tm} method to
obtain the solutions of the related Schr\"odinger equation in arbitrary $D$. Section IV
summarizes the results.

\section{Decomposition of the Kalb-Ramond Field}

The action for the Kalb-Ramond field in the de Sitter spacetime is given by
\begin{equation}
S=-\int d^{D}x\sqrt{-g}g^{\mu\nu}g^{\alpha\beta}g^{\gamma\theta}H_{\mu\alpha\gamma}H_{\nu\beta\theta},
\end{equation}
where $g_{\mu\nu} =(-1,e^{2H_{0}t},e^{2H_{0}t},e^{2H_{0}t})$ is the metric and $H_{\mu\nu\alpha}=\partial_ {[\mu} B_{\nu\alpha]}$. 
Differently from the procedure described in ref. \cite{Alencar:2011an}, here we will perform the field decomposition directly
from the equations of motion. From the above action we obtain the equation of motion
\begin{equation}
\partial_{\nu}(\sqrt{-g}g^{\mu\nu}g^{\alpha\beta}g^{\gamma\theta}H_{\mu\alpha\gamma})=0.
\end{equation}

Due to the gauge invariance we can impose the conditions $B_{i0}=0$ ($i=1,2...,D$) and due to the transversality condition we have $\partial^{i}B_{ij}=0$. Therefore, the only non trivial equations are
\begin{equation}\label{Bequation1}
\ddot{B}_{ij}+(D-5)\frac{\dot{a}}{a}\dot{B}_{ij}-\frac{1}{a^{2}}\nabla^{2}B_{ij}=0.
\end{equation}

Now we decompose the field as
\begin{equation}\label{Bequation2}
B_{ij}(\overrightarrow{x},t)=\sum_{\lambda}\int\frac{d^{D-1}k}{(2\pi)^{D-1}}\epsilon_{ij}^{\lambda}(u_{1}^{(\lambda)}(t)e^{i\overrightarrow{k}\cdot\overrightarrow{x}}+u_{2}^{(\lambda)}(t)e^{-i\overrightarrow{k}\cdot\overrightarrow{x}}),
\end{equation}
where $\epsilon_{ij}$ represents the polarization that obey the Gauge
condition $k_{i}\epsilon_{ij}=0$ and $\lambda$ run from $1$ to
$(D-2)(D-3)/2$. By using Eqs. (\ref{Bequation1}) and (\ref{Bequation2}) we get the equation
of motion for the modes $u$ given by 
\begin{equation}
\ddot{u}_{\sigma}+(D-5)\frac{\dot{a}}{a}\dot{u}_{\sigma}+\frac{1}{a^{2}}k^{2}u_{\sigma}=0\label{mode equantion},
\end{equation}
where $\sigma$ represents all the index of $u$, namely $u_{1}^{\lambda}$
or $u_{2}^{\lambda}$. 

Consider now the Hamiltonian of the harmonic
oscillator with time-dependent mass and frequency given by
\begin{equation}\label{hamiltonian}
H(t)=\frac{p^{2}}{2m(t)}+\frac{1}{2}m(t)\omega^{2}(t)q^{2}.
\end{equation}

The equation of motion is given by
\begin{equation}
\ddot{q}+\frac{\dot{m}}{m}\dot{q}+\omega^{2}q=0\label{HO},
\end{equation}
which is very simillar to Eq. (\ref{mode equantion}). Therefore
our system can be considered as a time-dependent harmonic oscillator
if we use $m(t)=a^{(D-5)}$ and $\omega=ka^{-1}$. A very
powerfull tool to perform the quantization of this system is discussed in the next section.

\section{Quantization of the Kalb-Ramond Field in the de Sitter Spacetime}

Consider a time-dependent harmonic oscillator described by Eq.
(\ref{hamiltonian}). It is well known that an invariant for Eq. (\ref{hamiltonian})
is given by \cite{Carinena}
\begin{equation}
I=\frac{1}{2}\left[\left(\frac{q}{\rho}\right)^{2}+(\rho p-m\dot{\rho}q)^{2}\right]\label{invariantedef},
\end{equation}
where $\rho(t)$ satisfies the generalized Milne-Pinney \cite{Milne,Pinney} (MP) equation 
\begin{equation}
\ddot{\rho}+\gamma(t)\dot{\rho}+\omega^{2}(t)\rho=\frac{1}{m^{2}(t)\rho^{3}}\label{MP},
\end{equation}
and $\gamma(t)=\dot{m}(t)/m(t)$. The invariant $I(t)$ satisfies
the equation
\begin{equation}
\frac{dI}{dt}=\frac{\partial I}{\partial t}+\frac{1}{i\hbar}[I,H]=0,
\end{equation}
and can be considered hermitian if we choose only the real solutions
of Eq. (\ref{MP}). Its eingenfunctions, $\phi_{n}(q,t)$, are assumed
to form a complete orthonormal set with time-independent discrete
eigenvalues, $\lambda_{n}=(n+\frac{1}{2})\hbar$.

Consider the time-dependent creation $(a^{\dagger}(t))$ and annihilation $(a(t))$ operators
defined as
\begin{eqnarray}
a^{\dagger}(t)&=&\left(\frac{1}{2\hbar}\right)^{1/2}\left[\left(\frac{q}{\rho}-i(\rho p-m(t)\dot{\rho}q)\right)\right];
\\
a(t)&=&\left(\frac{1}{2\hbar}\right)^{1/2}\left[\left(\frac{q}{\rho}+i(\rho p-m(t)\dot{\rho}q)\right)\right],
\end{eqnarray}
where $[a^{\dagger}(t),a(t)]=1$. In terms of $a(t)$ and $a^{\dagger}(t)$ the invariant $I$(see Eq. \ref{invariantedef}) can be written as
\begin{equation}
 I=\hbar \left(a^{\dagger}(t)a(t)+\frac{1}{2}\right).
\end{equation}

Let $|n,t>$ be the eigenstates of $I$. Therefore the following relations hold
\begin{eqnarray}
 a(t)|n,t>&=&\sqrt{n}|n-1,t>;
\\
 a^{\dagger}(t)|n,t>&=&\sqrt{n+1}|n+1,t>;
\\
I|n,t>&=&(n+\frac{1}{2})|n,t>.
\end{eqnarray}

The exact solution of the Schr\"odinger equation for the time-dependent harmonic
oscillator reads
\begin{equation}
|\psi_{n}>=e^{i\theta_{n}(t)}|n,t>,
\end{equation}
where the phase functions $\theta_{n}(t)$ satisfy the equation
\begin{equation}
\hbar\frac{d\theta_{n}(t)}{dt}=\left\langle \phi_{n}(q,t)\left|i\hbar\frac{\partial}{\partial t}-H(t)\right|\phi_{n}(q,t)\right\rangle.
\end{equation}

Since $I(t)$ expressed in terms of $a^{\dagger}(t)$ and $a(t)$ follows the same algebra as the hamiltonian for the time-dependent harmonic oscillator, it is not difficult to show that in the coordinate space $\psi_{n}(q,t)$ reads
\begin{eqnarray}
\psi_{n}(q,t)= & e^{i\theta_{n}(t)}\left(\frac{1}{\pi^{1/2}\hbar^{1/2}n!2^{n}\rho}\right)^{1/2}\times\nonumber \\
 & \exp\left\lbrace \frac{im(t)}{2\hbar}\left[\frac{\dot{\rho}}{2\hbar}+\frac{i}{m(t)\rho^{2}(t)}\right]q^{2}\right\rbrace \times\label{psi}\\
 & H_{n}\left(\frac{1}{\sqrt{\hbar}}\frac{q}{\rho}\right),\nonumber 
\end{eqnarray}
where $H_{n}$ is the Hermite polynomial of order $n$. 
For a given $m(t)$ and $\omega(t)$ one has to solve Eq. (\ref{MP}) to obtain the exact solutions to
the time-dependent Schr\"odinger equation for $H(t)$ given by Eq. (\ref{hamiltonian}). With $m(t)=a^{D-5}(t)$ 
and $\omega(t)=ka^{-1}(t)$ the MP equation reads
\begin{equation}
\ddot{\rho}+(D-5)\frac{\dot{a}}{a}\dot{\rho}+k^{2}a^{-2}(t)\rho=\frac{a^{-2(D-5)}(t)}{\rho^{3}}.\label{classicaleq}.
\end{equation}

According to Ref. \cite{Prince} the solution of the MP equation is related to the solutions
of Eq. (\ref{mode equantion}). By setting $dt=a(t)d\eta$ and $u_{\sigma}=\Omega\bar{u}_{\sigma}$, we rewrite Eq. (\ref{mode equantion}) as
\begin{eqnarray}
 &  & \bar{u}_{\sigma}''+\left(2a\frac{\dot{\Omega}}{\Omega}-\dot{a}+(D-5)\dot{a}\right)\bar{u}_{\sigma}'\nonumber \\
 &  & +\left(k^{2}+a^{2}\frac{\ddot{\Omega}}{\Omega}+(D-5)a\dot{a}\frac{\dot{\Omega}}{\Omega})\right)\bar{u}_{\sigma}=0,\label{classicaleqD}
\end{eqnarray}
where the prime and the dot denote the differentiation with respect to $\eta$ and $t$ respectively. By choosing $\Omega=a^{-\frac{D-5}{2}}$
we find 
\begin{eqnarray}
 &  & \bar{u}_{\sigma}''-\dot{a}\bar{u}_{\sigma}'+[k^{2}-\frac{(D-5)}{2}a\ddot{a}\nonumber \\
 &  & +\frac{(D-5)(D-7)}{4}\dot{a}^{2}]\bar{u}_{\sigma}=0.
\end{eqnarray}

Next, let us consider the de Sitter spacetime where $a=e^{Ht}$, and
with $dt=a(t)d\eta$ we get the relations 
\begin{equation}
\eta=-\frac{e^{-Ht}}{H}=\frac{1}{Ha(t)},\;\dot{a}=-\frac{1}{\eta},\;\ddot{a}=-\frac{H}{\eta},
\end{equation}
to obtain 
\begin{equation}
\bar{u}_{\sigma}''+\frac{1}{\eta}\bar{u}_{\sigma}'+(k^{2}-\frac{(D-5)^{2}}{4\eta^{2}})\bar{u}_{\sigma}=0,
\end{equation}
or
\begin{equation}
\left[\frac{d^{2}}{d(k\eta)^{2}}+\frac{1}{(k\eta)}\frac{d}{d(k\eta)}+(1-\frac{\nu^{2}}{(k\eta)^{2}})\right]\bar{u}_{\sigma}=0,
\end{equation}
where $\nu^{2}=\frac{(D-5)^{2}}{4}$. This is a Bessel equation  whose two independent 
solutions are $J_{\nu}(k|\eta|)$ and $N_{\nu}(k|\eta|)$. In terms of $u_{\sigma}$, we find
\[
\begin{cases}
(a^{\frac{-(D-5)}{2}}J_{\nu}(k|\eta|)\\
a^{\frac{-(D-5)}{2}}N_{\nu}(k|\eta|)
\end{cases},
\]
and the solution of the MP equation reads
\begin{eqnarray}
\rho=\frac{a^{\frac{-(D-5)}{2}}H2^{\frac{1}{2}}}{\pi}\left[AJ_{\nu}^{2}+BN_{\nu}^{2}+(AB-\frac{\pi^{2}}{4H^{2}})^{\frac{1}{2}}J_{\nu}N_{\mu}\right]^{\frac{1}{2}},
\end{eqnarray}
where $A$ and $B$ are real constants. The values of these constants are 
related to the choice of the vacuum. By  choosing the Bunch-Davies vacuum, which is the adiabatic vacuum at early
times we set $A=B=\pi/2H$ and $\rho$ reads, for $a=1/H\eta$
\begin{eqnarray}
\rho=(|H|\eta)^{\frac{(D-5)}{2}}\sqrt{\frac{H}{\pi}}\left[J_{\nu}^{2}+N_{\nu}^{2}\right]^{\frac{1}{2}}.\label{solution}
\end{eqnarray}

Since the problem of the KR field quantization in the de Sitter background was
reduced to solve the Schr\"odinger equation for the harmonic oscillator with time
dependent mass and frequency, the quantization procedure ends by substituting Eq. (\ref{solution})
into Eq. (\ref{psi}).

\section{Concluding Remarks}
In this paper we used the Lewis and Riesenfeld method to obtain the
time-dependent Schr\"odinger states emerging from the quantization of
the Kalb-Ramond field in the $D$-dimensional de Sitter spacetime.
It is well-known that a challenge in obtaining the exact solution
(see Eq. (\ref{psi})) for the SE with $H$ given in Eq. (\ref{hamiltonian}),
is to solve the Milne-Pinney (in terms of the universal scale
factor of universe $a(t)$) equation (\ref{classicaleqD}). 

In a previous article \cite{Alencar:2011an} we found that for $D=4$ the solution for the eletromagnetic field is time
independent. This is related to the fact the in $D=4$ the
Electromagnetic field is conformal and therefore comoving referential
do not feel a Bunch-Davies thermal bath. In the case of the KR field,
the solution (\ref{solution}) show us that for $D=6$ we have a time
independent solution,i.e. the KR field is conformal only in $D=6$. Therefore a solution with
arbitrary $a(t)$ can be found trivially in the conformal time. However
if we consider the space with dimensions $D\neq 6$, the KR
field looses its conformality. Therefore
this may have important consequences depending on the model considered.

For $D\neq6$ we can easily verify that $\rho=constant$ is not a
solution of the Eq. (\ref{solution}). In the case of a de Sitter spacetime,
we found the general solution and as we expected, the KR field has
more interesting solutions. For $D\neq6$ the solution are time-dependent,
what give us particle production and the comoving referential in this
spaces should feel a thermal bath. The drawback in this analysis is
that it is very dificult to imagine a way of measuring the associated
temperature of the Kalb-Ramond field. However, in interacting models
where the KR field is coupled to the Gauge field, some mesurable effects
may arise. For instance the Cremmer-Scherk-Kalb-Ramond \cite{Cremmer:1973mg}. In this case
we could have some correction to the Cosmic Microwave Background. 

\section*{Acknowledgments}

We acknowledge the financial support provided by Funda\c c\~ao Cearense de Apoio ao Desenvolvimento Cient\'\i fico e Tecnol\'ogico (FUNCAP), the Conselho Nacional de 
Desenvolvimento Cient\'\i fico e Tecnol\'ogico (CNPq) and FUNCAP/CNPq/PRONEX.

\end{document}